\documentclass[manuscript]{aastex}
\usepackage{spr-astr-addons}

\shorttitle{Wormholes}

\shortauthors{Rahaman et al.}

\begin{document}

\title{Wormholes supported by two non-interacting fluids}

\author{Farook Rahaman\altaffilmark{1}}
\altaffiltext{1}{Department of Mathematics, Jadavpur University,
Kolkata 700 032, West Bengal, India\\ rahaman@iucaa.ernet.in}

\author{Saibal Ray\altaffilmark{2}}
\altaffiltext{2}{Department of Physics, Government College of
Engineering \& Ceramic Technology, Kolkata 700 010, West Bengal,
India\\ saibal@iucaa.ernet.in}

\author{Safiqul Islam\altaffilmark{3}}
\altaffiltext{3}{Department of Mathematics, Jadavpur University,
Kolkata 700 032, West Bengal, India\\ sofiqul001@yahoo.co.in}

\begin{abstract}
We provide a new matter source that supplies fuel to construct
wormhole spacetime. The exact wormhole solutions are found in the
model having, besides real matter, an anisotropic dark energy. We
have shown that the exotic matters that are the necessary
ingredients for wormhole physics violate null and  weak energy
conditions  but obey strong energy condition marginally. Though
the wormhole comprises of exotic matters yet the effective mass
remains positive. We have calculated the effective mass of the
wormhole up to $8$~km from the throat (assuming throat radius as
$4$~km) as $1.3559M_\odot $. Some physical features are briefly
discussed.
\end{abstract}


\section{Introduction}
It was revealed by the observations on supernova due to the
High-$z$ Supernova Search Team (HZT) and the Supernova Cosmology
Project (SCP) \citep{Riess1998,Perlmutter1998} that the present
expanding Universe is getting gradual acceleration. As a cause of
this acceleration it is argued that a kind of exotic matter having
repulsive force  is responsible for speeding up the Universe some
$7$ billion years ago. To understand the nature of this
hypothetical energy that tends to increase the rate of expansion
of the Universe several models have been proposed by the
scientists so far \citep{Overduin1998,Sahni2000}.

As far as matter content of the Universe is concerned, it is
convincingly inferred from distant supernovae, large scale
structure and CMB, that $96\%$ of matter is hidden mass
constituted by $23\%$ dark matter and $73\%$ unknown exotic matter
known as dark energy whereas only $4\%$ mass in the form of
ordinary mass which is visible contrary to the non-luminous dark
matter \citep{Pretzl,Freeman,Wheeler,Gribbin}.

On the other hand, theoretically a {\it wormhole}, which is
similar to a tunnel with two ends each in separate points in
spacetime or two connecting black holes, was conjectured first by
Weyl \citep{Coleman1985} and later on by \citet{Wheeler1957}. This
is essentially some kind of hypothetical topological feature of
spacetime which may acts as {\it shortcut} through spacetime. In
principle this means that a wormhole would allow travel in time as
well as in space and can be shown explicitly how to convert a
wormhole traversing space into one traversing time
\citep{Morris1988a}. The possibility of traversable wormholes in
general relativity was demonstrated by \citet{Morris1988b} which
held open by a spherical shell of exotic matter whereas quite a
number of wormhole solutions were obtained much earlier with
different physical motivation by other scientists
\citep{Ellis1973,Bronnikov1973,Clement1984}.

 However, other types of wormholes where the traversing
path does not pass through a region of exotic matter were also
available in the literature \citep{Visser1989,Visser1995}.

In this connection we are interested to mention that in some of
our previous works we dealt with a new type of thin-shell wormhole
constructed by applying the cut-and-paste technique to two copies
of a charged black hole \citep{Usmani2010}. This has been done in
generalized dilaton-axion gravity which was inspired by low-energy
string theory. This was done following the work of
\citet{Visser1989}, who proposed a theoretical method for
constructing a new class of traversable Lorentzian wormholes from
black-hole spacetimes by surgical grafting of two Schwarzschild
spacetimes. The main benefit in Visser's approach is that it
minimizes the amount of exotic matter required.

However, the necessary ingredients that supply fuel to construct
wormholes remain an elusive goal for theoretical physicists.
Several proposals have been proposed in literature
\citep{kuh1999,sus2005,lobo2005b,lobo2005d,zas2005,das2005,rahaman2006,rahaman2007,rahaman2008,rahaman2009a,rahaman2009b,kuh2010,jamil2010}.
In the present work taking cosmic fluid as source we have provided
a new class of wormhole solutions under the framework of general
relativity. Here this matter source would supply fuel to construct
the exact wormhole spacetime. Besides the real matter source an
anisotropic dark energy also considered here. Regarding anisotropy
of dark energy we notice that several works are now available in
the literature \citep{Richard,Campanelli,Stephen} which support
this idea.

It is shown in the present investigation that the exotic matters
violate null and weak energy conditions but obey strong energy
condition marginally. The wormhole constructed here in the
presence of real and exotic matters provides a positive effective
mass. This effective mass of the wormhole is $1.3559M_\odot $ up
to $4$ km throat radius. The plan of the investigation is as
follows: in Sec. 2 basic equations for constructing wormhole are
provided and as a result some toy models for wormholes are
presented in Sec. 3 whereas in Sec. 4 we have discussed various
physical features of the model supported by exotic matters. In
Sec. 5 specific concluding remarks are made.

\section{Basic equations for constructing wormhole}
The metric for a static spherically symmetric spacetime is taken
as
\begin{equation}
ds^{2}=-e^{\nu(r)}dt^{2}+e^{\lambda(r)}dr^{2}+r^{2}(d\theta^{2}+sin^{2}\theta
d\phi^{2}), \label{Eq3}%
\end{equation}
where $r$ is the radial coordinate. Here $\nu$ and $\lambda$ are
the metric potentials which have functional dependence on $r$.

We propose matter sources, which constitutes with two
non-interacting fluids, as follows: the first one is real matter
in the form of perfect fluid and the second one is anisotropic
dark energy which is phantom energy type. The mining of this
second ingredient can be done from cosmic fluid that is
responsible for acceleration of the Universe \citep{Rahaman2012}.

Therefore, the energy-momentum tensors can be expressed in the
following form

\begin{equation}
\label{eq2} T^{0}_{0}=-\rho^{eff}\equiv -(\rho + \rho^{de})
\end{equation}
\begin{equation}
T^{1}_{1} = p_r^{eff}\equiv (p +p^{de}_r)
\end{equation}
\begin{equation}
T^{2}_{2}=T^{3}_{3}=p_t^{eff}\equiv (p+p^{de}_t),
\end{equation}
where $\rho^{de}$, $p^{de}_r$ and $p^{de}_t$ are dark energy
density, dark energy radial pressure and dark energy transverse
pressure respectively whereas $\rho$ and $p$ are assigned for the
real matter.

Now, we specially consider that the dark energy radial pressure is
proportional to the dark energy density, so that
\begin{equation}
\label{eq3} p^{de}_r=-\omega \rho^{de}, \quad \omega>1.
\end{equation}

Also, we assume that the dark energy density is proportional to
the mass density
\begin{equation}
\label{eq4} \rho^{de}=n\rho.
\end{equation}
Here the constraint to be imposed is $n > 0$.

In connection to the {\it ansatz (5)} it is worthwhile to mention
that the equation of state of this type which implies that the
matter distribution under consideration is in is phantom energy
type \citep{lobo2005d}. However, for $\omega = 1$, the matter
distribution   is known as a `false vacuum' or `degenerate vacuum'
or `$\rho$-vacuum'
\citep{Blome1984,Davies1984,Hogan1984,Kaiser1984}.

Now, as usual we employ use the following standard equation of
state (EOS)
\begin{equation}p = m \rho, \quad 0<m<1,
\end{equation}
where $m$ is a parameter corresponding to normal matter. The
Einstein equations are
\begin{equation}
\label{eq5}
e^{-\lambda}\left(\frac{\lambda^\prime}{r}-\frac{1}{r^2}\right) +
\frac{1}{r^2} = 8\pi\left(\rho+\rho^ {de}\right),
\end{equation}
\begin{equation}
\label{eq6}
e^{-\lambda}\left(\frac{\nu^\prime}{r}+\frac{1}{r^2}\right) -
\frac{1}{r^2} = 8\pi\left(p+p^{de}_r\right),
\end{equation}
\begin{equation}
\label{eq7} \frac{1}{2}e^{-\lambda}\left[\frac{1}{2}{\nu^\prime}^2
+ \nu^{\prime\prime}-\frac{1}{2}\lambda^{\prime}\nu^\prime
+\frac{1}{r}(\nu^\prime-\lambda^\prime)\right] =
8\pi\left(p+p^{de}_t\right).
\end{equation}
\label{eq8}

The generalized Tolman-Oppenheimer-Volkov (TOV) equation is
\begin{equation}
\label{eq13} \frac{d(p_r^{eff})}{dr}   +
\frac{\nu^\prime}{2}\left(\rho^{eff} +p_r^
{eff}\right) +
\frac{2}{r}\left(p_r^{eff} - p_t^{eff}\right) = 0.
\end{equation}

Let us write the metric coefficient $g_{rr}$ as
\begin{equation}
\label{eq9} e^{-\lambda(r)}=1 - \frac{ b(r)}{r},
\end{equation}
where, $b(r)$ is the shape function of the wormhole structure
which can easily be recognized as mass function
\citep{Landau1959}.

Here, the above shape function, by the use of the Eqs. (6) and
(8), can be expressed as
\begin{equation} b(r)=  8\pi \int(\rho + \rho^{de}) r^2 dr
= 8\pi \int \rho(1 + n) r^2 dr.
\end{equation}

From the field Eqs. (8) and (9), via the ansatz (5), we get
\begin{equation}
\label{eq10} 8\pi(\rho + \rho^{de} +p+p_r^{de}) =
e^{-\lambda}\left(\frac{\lambda^\prime}{r} +
\frac{\nu^\prime}{r}\right).
\end{equation}
which readily gives
\begin{equation} \label{eq9} \nu = \int e^\lambda  [~ 8 \pi \rho(1+m+n-n \omega) r
+(e^{-\lambda})^\prime~]dr.
\end{equation}

\section{Toy models for wormholes}
Now we consider several toy models for the present case of
wormholes.

\subsection{Specific shape function}
Consider the specific form of shape function as
\begin{equation} b(r) =  r_0\left(\frac{r}{r_0}\right)^{\alpha},
            \label{Eq1}
\end{equation}
where $r_0$ corresponds to the wormhole throat and $\alpha$ is an
arbitrary constant.

Using the above shape function (16) in the field equations, we get
the following expressions of the parameters
\begin{equation}
\rho =
\frac{\alpha}{8\pi(1+n)r_0^2}\left(\frac{r}{r_0}\right)^{\alpha-3},
            \label{Eq1}
\end{equation}
\begin{equation}
\nu =  1-\frac{A}{(\alpha-1)} \ln \left[ 1 -
       \left(\frac{r}{r_0}\right)^{\alpha-1}\right],
            \label{Eq1}
\end{equation}

\begin{equation}
p= m \rho =   \frac{m \alpha
          }{8\pi(1+n)r_0^2}\left(\frac{r}{r_0}\right)^{\alpha-3},
            \label{Eq1}
\end{equation}

\begin{equation}
p^{de}_r = - \omega \rho^{de} = - \omega n \rho  =  - \frac {
\omega n \alpha
}{8\pi(1+n)r_0^2}\left(\frac{r}{r_0}\right)^{\alpha-3},
            \label{Eq1}
\end{equation}

\begin{eqnarray}
p^{de}_t= \frac{\alpha (\alpha-3)(m-\omega
n)}{16\pi(1+n)r_0^2}\left(\frac{r}{r_0}\right)^{\alpha-3} -\frac{
\omega n \alpha
}{8\pi(1+n)r_0^2}\left(\frac{r}{r_0}\right)^{\alpha-3} \nonumber
\\ + \frac{\alpha (1+m)A}{32\pi(1+n)r_0^2\left[ 1-
\left(\frac{r}{r_0}\right)^{\alpha-1}\right]}\left(\frac{r}{r_0}\right)^{2\alpha-4},
\end{eqnarray}

where
\begin{equation}
A= \left[\frac{(1+m+n- n\omega)\alpha}{(1+n)} +(1-\alpha)\right].
\end{equation}

Since the spacetime is asymptotically flat, we demand integration
constant to be unity.

One can note that, $\frac{b(r)}{r} ~\longrightarrow 0$ as $r
\longrightarrow \infty$ implies $\alpha<1$. Also, flare-out
condition, which can be found out by taking the derivative of the
shape function $b(r)$ at $r=r_0$ i.e. $b^\prime(r_0) < 1$ gives,
$\alpha<1$.

\subsection{Specific energy density}
Let us consider the energy density function as
\begin{equation}
\rho(r) = \rho_0\left(\frac{r_0}{r}\right)^\beta.
\end{equation}
Here, $r_0$ is the wormhole throat and $\rho_0 >0$ corresponds to
the energy density at the throat and  $\beta$ is an arbitrary
constant.

Using the above energy density function (23), one can get the
solutions of the parameters characterized the wormhole as
\begin{equation}
b(r) =  \frac{ 8 \pi (1+n)\rho_0 r_0^\beta
r^{3-\beta}}{(3-\beta)}.
\end{equation}
At the  throat radius $r=r_0$,  $b(r_0) = r_0$ and this implies
\begin{equation}\rho_0 = \frac{(3-\beta)}{8 \pi
(1+n)r_0^2}.
\end{equation}
Here,   $\rho_0 >0$ implies $\beta<3$.

 Using the value of $\rho_0$ in Eq. (25),
one gets the following form of the shape function as
\begin{equation}
b(r) =  r_0\left(\frac{r}{r_0}\right)^{3-\beta}.
\end{equation}

Now the other parameters can be found as
\begin{equation}
e^\nu = \left[ 1 -\left(\frac{r}{r_0}\right)^{2-\beta} \right]^B
\end{equation}
where
\begin{equation}
B=\left[ \frac{3-\beta}{(\beta-2)(1+n)}\right] \left[(m-n \omega)
+ \frac{1+n}{3-\beta}\right],
\end{equation}
 \begin{eqnarray}
p^{de}_t = \frac{\beta(3-\beta)(n \omega-m)}{16 \pi (1+n)r_0^2}
\left(\frac{r}{r_0}\right)^{-\beta} -\frac{n \omega (3-\beta)}{8
\pi (1+n)r_0^2} \left(\frac{r}{r_0}\right)^{-\beta} \nonumber \\ -
\frac{B(3-\beta)(2-\beta)(1+m)}{32 \pi (1+n)r_0^2 \left[ 1 -
\left(\frac{r}{r_0}\right)^{2-\beta}\right]}
\left(\frac{r}{r_0}\right)^{2-2\beta}.
\end{eqnarray}
One can note that $\frac{b(r)}{r} ~\longrightarrow 0 $ as
$r\longrightarrow\infty $ implies $\beta>2$. Also, flare-out
condition $b^\prime(r_0) < 1$ gives $\beta>2$. Therefore the
possible rang of $\beta$ is $2<\beta<3$.

\subsection{Constant redshift function}
Consider the constant redshift function and without loss of
generality we assume
\begin{equation}
\nu =  0.
            \label{Eq1}
\end{equation}

Here all the parameters are
\begin{equation}
b(r) =   \left(\frac {r}{\gamma_0}\right)^{\gamma},
            \label{Eq1}
\end{equation}
where $\gamma =\frac{1+n}{n \omega -m}$ and $\gamma_0$ is an
integration constant. Note that, at the throat radius $r=r_0$,
$b(r_0) = r_0$ implies $\gamma_0 = r_0^{\frac{\gamma-1}{\gamma}}$.
Thus $b$ takes the form as
\begin{equation}
b(r) =  r_0 \left(\frac {r}{r_0}\right)^{\gamma}.
            \label{Eq1}
\end{equation}

The other parameters are
\begin{equation}
\rho =  \frac{\gamma}{8\pi(1+n)r_0^2}\left(\frac
{r}{r_0}\right)^{\gamma-3},
            \label{Eq1}
\end{equation}

\begin{equation}
p= \frac{m \gamma}{8\pi(1+n)r_0^2}\left(\frac
{r}{r_0}\right)^{\gamma-3},
            \label{Eq1}
\end{equation}

\begin{equation}
p^{de}_r = - \omega \rho^{de} = -n \omega  \rho  = - \frac{n
\omega  \gamma}{8\pi(1+n)r_0^2}\left(\frac
{r}{r_0}\right)^{\gamma-3},
            \label{Eq1}
\end{equation}

\begin{equation}
p^{de}_t = \left[ \frac{\gamma(m - n \omega)(\gamma-3)-2n \gamma
\omega }{16\pi(1+n)r_0^2}\right]\left(\frac
{r}{r_0}\right)^{\gamma-3}.
            \label{Eq1}
\end{equation}

One can note that, if one chooses  the values  parameter $n$, $m$,
$\omega$ for which $\gamma
> 1$, then  $\frac{b(r)}{r}$ does not tend to zero as $r
\longrightarrow \infty$. This implies that the solution is not
asymptotically flat. So, we have to match our interior solution to
the exterior Schwarzschild solution. According to
\citet{Morris1988a,Morris1988b} for traversable wormhole the
spacetime is to be nearly flat i.e. $\frac{b(a)}{a} << 1$ for cut
off at some $r=a$. Unfortunately, since $\gamma>1$, we can not get
$a>r_0$, for which $\frac{b(a)}{a} << 1$. Thus $\gamma
> 1$ is not acceptable. However, $\gamma
< 1$  implies $\frac{b(r)}{r}$ tends to zero as $r \longrightarrow
\infty$. Note that one can never choose $n \omega =m $.

\section{Some features of the models}

\subsection{Visual Structure}
Fortunately, all the three models have the shape functions that
are of polynomial form of different power index i.e. $b(r) =
 r_0\left(\frac{r}{r_0}\right)^X$, where,\\
   $X = \alpha$,  for model-I,\\
      $~~~$ = $3-\beta $, for~~model-II,\\
       $~~~$ = $\gamma $, for~~model-III.\\

We note that for $A \ne 0$ in Eq. (18) and  $B \ne 0$ in Eq. (27),
$g_{tt} = 0$ at $r = r_0$. This indicates that there is an
infinite redshift at  $r=r_0$  and  the system is not a wormhole.
This $r=r_0$ is either a black hole horizon or a singularity. In
other words, these solutions reflect a non-traversable wormholes.
However, if we impose the conditions $A=0$ in Eq. (18) and $B=0$
in Eq. (27), then for both cases, one gets $e^\nu =1$ (re-scaling
the case given in Subsection {\bf 3.1}) and rendering them
traversable.

Now, the conditions $A=0$ and $B=0$ imply,
\begin{equation}
X= \alpha = 3-\beta = \gamma= \frac{1+n}{n \omega -m}.
\end{equation}
As discussed in Sec. {\bf 3.3}, we should choose the value of X
less that unity.

According to \citet{Morris1988a,Morris1988b}, one can picture the
special shape of the wormhole  by rotating the profile curve
$z=z(r)$ about the $z-$axis. This curve is defined by
\begin{equation}
\frac{dz}{dr}=\pm \frac{1}{\sqrt{\displaystyle{r/b(r)}-1}}= \pm
\frac{1}{\sqrt{\left(\frac{r}{r_0}\right)^{1-X}-1}}.
\end{equation}
One can note from the definition of wormhole that at $r= r_0$ (the
wormhole throat) Eq. (38) is divergent i.e. embedded surface is
vertical there.

For the specific value of X, say $X=0.5$, we draw the embedded
diagram of the wormhole which is shown in Fig. 1. One can note
that this value of X can be achieved by choosing $\alpha=0.5$ in
model-I, $\beta = 2.5$ in model-II and  $\omega=5$, $m=0.4$ and
$n=0.8$ in model-III.

The surface of revolution of the curve about vertical z axis makes
the diagram complete. The full visualization of the surface
generated by the rotation of the embedded curve about the vertical
z axis is shown in Fig. 2.

\begin{figure}[htbp]
\centering
\includegraphics[scale=.3]{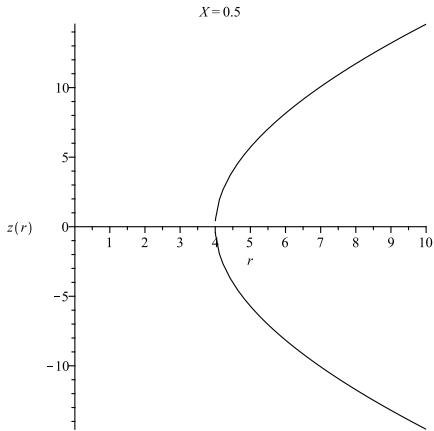}
\caption{The profile curve of the wormhole.\label{fig1}}
\end{figure}

\begin{figure}[htbp]
\centering
\includegraphics[scale=.5]{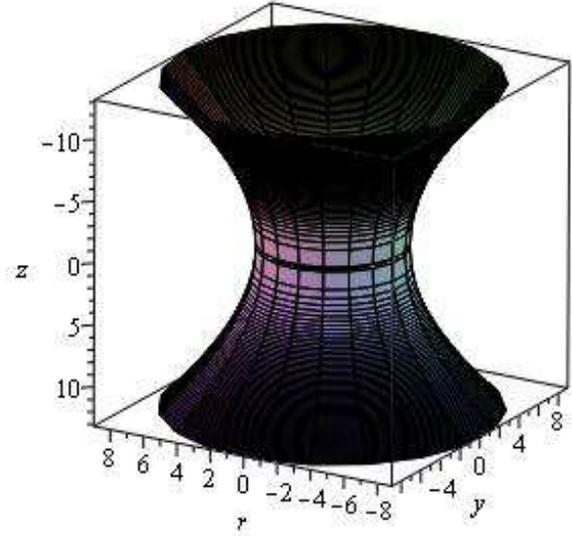}
\caption{The embedding diagram generated by rotating
    the profile curve about the $z-$axis.\label{fig2}}
\end{figure}

According to \citet{Morris1988b}, the $r$-coordinate is
ill-behaved near the throat, but proper radial distance
\begin{equation}
 l(r) = \pm \int_{r_0^+}^r \frac{dr}{\sqrt{1-\frac{b(r)}{r}}}
            \label{Eq20}
          \end{equation}
must be well behaved everywhere i.e. we must require that $l(r)$
is finite throughout the spacetime.

The proper radial distance $l(r)$ from the throat to a point
outside is given in Fig. 3.

\begin{figure}[htbp]
\centering
\includegraphics[scale=.3]{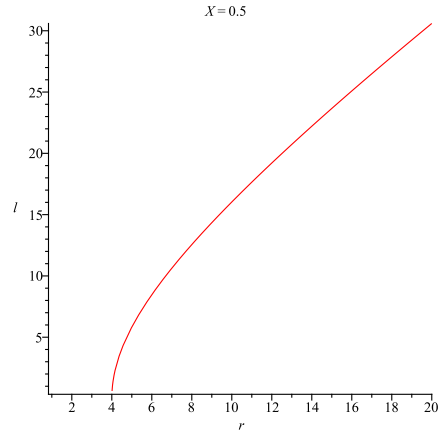}
\caption{The graph of the radial proper distance
$l(r)$\label{fig3}}
\end{figure}

\subsection{Energy Conditions}
Now, we check the material compositions comprising the wormhole
whether it will satisfy or not the null energy condition (NEC),
weak energy condition (WEC) and strong energy condition (SEC)
simultaneously at all points outside the source. Since we write
all equations in terms of $X$ and follow the assumptions $A=0$ and
$B=0$, we have
\begin{equation}
\rho^{eff} =  \frac{X}{8 \pi r_0^2}
\left(\frac{r}{r_0}\right)^{X-3},
\end{equation}
\begin{equation} p_r^{eff} =
-\frac{1}{8 \pi r_0^2} \left(\frac{r}{r_0}\right)^{X-3},
\end{equation}
\begin{equation}
p_t^{eff} = \frac{(1-X)}{16 \pi r_0^2}
\left(\frac{r}{r_0}\right)^{X-3},
\end{equation}
\begin{equation}\rho^{eff}+p_r^{eff} =  \frac{(X-1)}{8 \pi r_0^2}
\left(\frac{r}{r_0}\right)^{X-3},\label{ec40}
\end{equation}
\begin{equation}
\rho^{eff} + p_t^{eff} = \frac{(1+X)}{16 \pi r_0^2}
\left(\frac{r}{r_0}\right)^{X-3},
\end{equation}
\begin{equation}
\rho^{eff} +p_r^{eff}+ 2p_t^{eff} = 0.
\end{equation}

The Fig. 4 indicates that  the null energy condition (NEC), weak
energy condition (WEC) are violated, however, the strong energy
condition (SEC) is satisfied marginally. Hence, in our models, the
null energy condition (NEC) is violated to hold a wormhole open.

\begin{figure}[htbp]
\centering
\includegraphics[scale=.4]{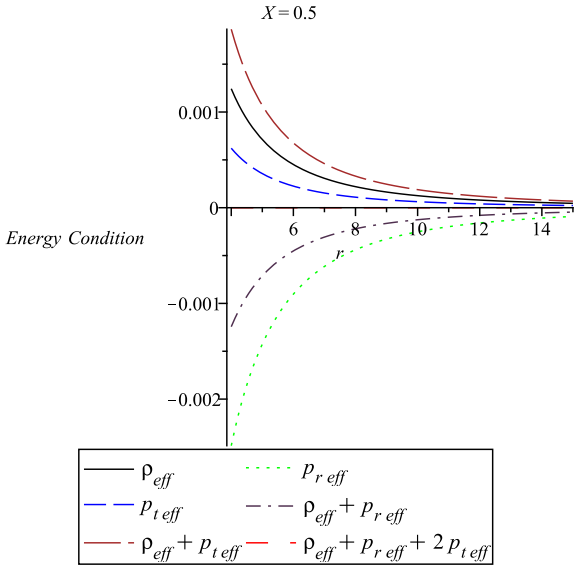}
\caption{The variation of left hand side of the expressions of
energy conditions are shown against $r$.\label{fig4}}
\end{figure}

\subsection{Equilibrium condition}
Following Ponce de Le\'{o}n \cite{Leon1993}, we write the TOV Eq.
(11) for an anisotropic fluid distribution, in the following form
\begin{equation}
-\frac{M_G\left(\rho^{eff}+p_r^{eff}\right)}{r^2}e^{\frac{\lambda-\nu}{2}}-\frac{dp_r^{eff}}{dr}
+\frac{2}{r}\left(p_t^{eff}-p_r^{eff}\right)=0, \label{eq31}
\end{equation}
where $M_G=M_G(r)$ is the effective gravitational mass within the
radius $r$ and is given by
\begin{equation}
M_G(r)=\frac{1}{2}r^2e^{\frac{\nu-\lambda}{2}}\nu^{\prime},\label{eq32}
\end{equation}
which can easily be derived from the Tolman-Whittaker formula and
the Einstein's field equations. Obviously, the modified TOV
equation (46) describes the equilibrium condition for the wormhole
subject to gravitational ($F_g$) and hydrostatic ($F_h$) plus
another force due to the anisotropic nature ($F_a$) of the matter
comprising the wormhole. Therefore, for equilibrium the above Eq.
(46) can be written as
\begin{equation}
 F_g+ F_h + F_a=0,\label{eq33}
\end{equation}
where,
\begin{equation}
F_g =-\frac{\nu^\prime}{2}\left(\rho^{eff} +p_r^{eff}\right) =
0,\label{eq34}
\end{equation}
\begin{equation}
F_h =-\frac{dp_r^{eff}}{dr}=   \frac{(X-3)}{8 \pi r_0^3}
\left(\frac{r}{r_0}\right)^{X-4},\label{eq35}
\end{equation}
\begin{equation}
F_a=\frac{2}{r}\left(p_t^{eff} -p_r^{eff}\right)\\ =
 \frac{(3-X)}{8 \pi r_0^3}
\left(\frac{r}{r_0}\right)^{X-4}.
\end{equation}
The profiles of $F_g$, $F_h$ and $F_a$ for our chosen source are
shown in Fig. 5. The figure indicates that equilibrium stage can
be achieved due to the combined effect of pressure anisotropic,
gravitational and hydrostatic forces. It is to be distinctly noted
that by virtue of the Eq. (30), the gravitational force term in
Eq. (49) vanishes which is readily observed from the Fig. 5 as the
plot for $F_g$ coincides with the coordinate $r$. The other two
plots reside opposite to each other to make the system balanced.

\begin{figure}[htbp]
\centering
\includegraphics[scale=.3]{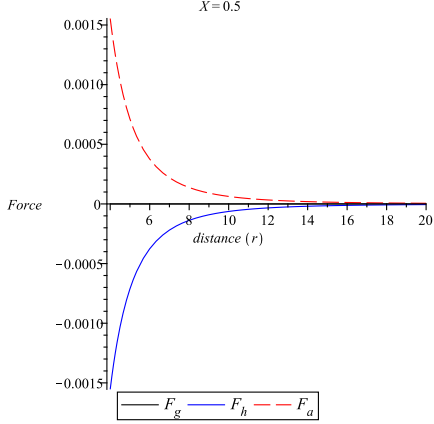}
\caption{Three different forces acting on fluid elements in static
equilibrium is shown against $r$.\label{fig5}}
\end{figure}

 \subsection{Effective gravitational mass}
In our model the effective gravitational mass,  in terms of the
effective energy density $\rho^{eff}$, can be expressed as
\begin{eqnarray}
\label{eq40}
M^{eff}=4\pi\int_{r_0}^{R}\left(\rho+\rho^{de}\right)r^2 dr
\nonumber \\ = 4\pi\int_{r_0}^{R}\left[\frac{X}{8 \pi r_0^2}
\left(\frac{r}{r_0}\right)^{X-3} \right]r^2 dr
 = \frac{R^X-r_0^X}{2r_0^{X-1}}.
\end{eqnarray}
   The effective mass of
the wormhole up to radius $8$~km from the throat (assuming the
throat radius $r_0 = 4$~km and X=0.5)  is obtained as $M^{eff} =
.828~km = 0.561~M_\odot $ (where $1$ Solar Mass $= 1.475$~km).

We note from the Eq. (52) that though wormholes are supported by
the exotic matter, but the effective mass is positive. This
implies that for an observer sitting at large distance could not
distinguish the gravitational nature between wormhole and a
compact mass $M$.

\subsection{Total gravitational energy}
It is known that total gravitational energy of a localized real
matter obeying all energy conditions is negative. Naturally, we
would like to know how the gravitational energy behaves for the
matters that supply fuel of our wormhole structure. Following
\citet{Lyndell2007} and \citet{Nandi2009}, we have the following
expression for the total gravitational energy of the wormhole as

\begin{equation}
\label{eq41} E_g= \frac{1}{2}  \int_{r_0}^r [ ~1 -\sqrt{g_{rr}}~]
\rho^{eff} r^2 dr +  \frac{r_0}{2},
\end{equation}
where the second part is the contribution from the effective
gravitational mass. It is to note that here the range of the
integration is considered from the throat $r_0$ to the embedded
radial space of the wormhole geometry. Here, the total
gravitational energy of the wormhole is given by

\begin{eqnarray}
\label{eq41} E_g=   \int_{r_0}^{r =a r_0} \left(\frac{X}{16 \pi
}\right)\left[ 1 -\sqrt{
\frac{1}{1-\left(\frac{r}{r_0}\right)^{X-1}}}\right]
\left(\frac{r}{r_0}\right)^{X-1} dr \nonumber\\+ \frac{r_0}{2}.
\end{eqnarray}

 For the specific value of X, say $X=0.5$, we calculate
 the numerical value of the integrand
(56) describing the total gravitational energy from the throat
$r_0 =4$ to the embedded radial space $1.5r_0 =7$ (i.e. $ a =1.5$)
as $ E_g =1.9397$, which indicates that $E_g > 0$, in other words,
there is a repulsion around the throat. This result is very much
expected for constructing a physically valid wormhole. It is to be
noted that the non-vanishing of $E_G$ explains why the wormhole is
able to affect on the test particles despite $g_{tt} $ = constant
\citep{Nandi2009}.

\subsection{Traversability conditions}
If the tidal gravitational forces felt by a traveler be reasonably
small, then travel through wormhole is possible. Due to
\citet{Morris1988a}, the acceleration experienced by the traveler
should be less than the Earth's gravity. A traveler of two meter
height feels  the tidal accelerations between two parts of his
body should be less than the gravitational acceleration at Earth's
surface $g_{earth}$ ($g_{earth}\approx 10 m /sec^2$). Now, the
testing tangential tidal constraint is given by (assuming $ \nu ^
\prime = 0$)

\begin{eqnarray}|R_{t \theta t \theta}| = R_{t \phi t \phi}| =
|\frac{\beta^2}{2r^2} (\frac{v}{c})^2 ( b^\prime - \frac{b}{r} ) |
\leq \frac{g_{earth}}{2c^2m } \nonumber\\ \approx \frac{1}{10^{10}
m^2}\end{eqnarray}

with $ \beta = \frac{1}{\sqrt{1-(\frac{v}{c})^2}}$ and c is the
velocity of light.

For $ v << c $, we have $\beta  \approx  1 $. We  substitute  the
expression of our shape function to yield a restriction for the
velocity as

\begin{equation} \frac{v}{c} < \frac{1}{10^8} \frac{2
r_0}{\sqrt{(1-X) (\frac{r}{r_0})^{X-2}}}\end{equation}

The above inequality represents the tangential tidal force and
restrict the speed v of the while crossing the wormhole. Here
radial acceleration is zero since $ R_{rtrt} = 0 $, for our
wormhole spacetime. Acceleration felt by a traveler should less
than the gravitational acceleration at earth surface, $g_{earth}$.
The condition imposed by \citet{Morris1988a} is as follows:

\begin{equation} |\textbf{f}| = |\sqrt{[ 1 - \frac{b(r)}{r}}] \beta
^ \prime c^2| \leq  g_{earth}
 ~~ for~~
\nu^\prime ~= ~0. \end{equation}

 For the traveler's velocity $ v = constant$, one finds that $|\textbf{f}|= 0$.
 In our model the the above condition  is automatically satisfied, the
 traveler feels a zero gravitational acceleration.

\section{Final Remarks}
In searching for a possibility of Lorentzian traversable wormhole
in general relativity we have, in the present paper, considered
the anisotropic dark energy along with the real matter source. The
novel point here seems to be the interpretation in terms of two
fluids, which is more or less arbitrary. We have constructed the
wormholes from three different points of view (namely, specific
shape function, specific  energy density  and constant redshift
function)  for the two non interacting fluids. To get realistic
models, one has to impose different restrictions on the
parameters. Fortunately, after imposing the restriction all the
three models give the same structure of the wormhole.

Our main observations of the present investigation are as follows:

(1) The exotic matter though as usual violates null and weak
energy conditions  but does obey strong energy condition
marginally.

(2) Since, $E_g > 0$, there is a repulsion around the throat which
is very much expected for valid construction of a wormhole.

Some of the other minor observations are as follows:

(1) For the spacetime to be asymptotically flat we note that,
$\frac{b(r)}{r} ~\longrightarrow 0$ as $r \longrightarrow \infty$
implies $X<1$. Flare-out condition, $b^\prime(r_0) < 1$ also
gives, $X<1$.

(2) To travel through a wormhole, the tidal gravitational forces
experienced by a traveler must be reasonably small. In our model
the the above condition is automatically satisfied, the traveler
feels a zero gravitational acceleration since  $\nu =0$.

Based on the above observations we would like to conclude that the
wormhole model provided here with anisotropic dark energy and real
matter  is fascinating in several aspects and hence very promising
one.

However, we observe in the present investigation that anisotropic
dark energy with different energy density and radial pressure may
also provide the exotic fuel in constructing the wormhole. So,
interpretations within dark energy or other than dark energy is
needed for exotic sector of the energy-momentum tensor which can
be sought for in a future work.

\section*{Acknowledgments} FR and SR are thankful to the authority of
Inter-University Centre for Astronomy and Astrophysics, Pune,
India for providing them Visiting Associateship under which a part
of this work was carried out. FR is also thankful to PURSE and UGC
for providing financial support. We are  also grateful to Prof. F.
N. Lobo and Dr. G. C. Shit for  several insightful comments on
this manuscript.

\end{document}